\documentclass[prx,aps,amsmath,amssymb,floatfix,twocolumn,longbibliography]{revtex4-1}

\pdfoutput=1
\usepackage{hyperref}
\usepackage{graphicx}
\usepackage[usenames]{color}
\usepackage{bm}

\def\eqq#1{Eq.~(\ref{#1})}

\def\eq#1{(\ref{#1})}

\def\f#1{Fig.~\ref{#1}}

\def\s#1{Section~\ref{#1}}

\def\c#1{~\cite{#1}}

\def\cc#1{Ref.~\cite{#1}}

\def\av#1{\left \langle #1 \right \rangle}

\def\beq{\begin{equation}}
\def\eeq{\end{equation}}
\def\bea{\begin{eqnarray}}
\def\eea{\end{eqnarray}}

\def\e{{\rm e}}
\def\tf{t_{\rm f}}

\def\llambda{{\bm \lambda}}
\def\xx{{\bm x}}
\def\ot{\leftarrow}

\definecolor{blue}{rgb}{0,0,0}
\newcommand{\bb}[1]{\textcolor{blue}{#1}}

\begin{document}

\title{Free-energy estimates from nonequilibrium trajectories under varying-temperature protocols}
\author{Stephen Whitelam}\email{swhitelam@lbl.gov}
\affiliation{Molecular Foundry, Lawrence Berkeley National Laboratory, 1 Cyclotron Road, Berkeley, CA 94720, USA}

\begin{abstract}
The Jarzynski equality allows the calculation of free-energy differences using values of work measured from nonequilibrium trajectories. The number of trajectories required to accurately estimate free-energy differences in this way grows sharply with the size of work fluctuations, motivating the search for protocols that perform desired transformations with minimum work. However, protocols of this nature can involve varying temperature, to which the Jarzynski equality does not apply. We derive a variant of the Jarzynski equality that applies to varying-temperature protocols, and show that it can have better convergence properties than the standard version of the equality. We derive this modified equality, and the associated fluctuation relation, within the framework of Markovian stochastic dynamics, complementing related derivations done within the framework of Hamiltonian dynamics.
\end{abstract}
\maketitle

\section{Introduction}
\label{introduction}

The Jarzynski equality allows the calculation of free-energy differences by measuring the work done by a nonequilibrium protocol. For a system at fixed temperature $\beta^{-1}$ that is initially in equilibrium and driven out of it by the variation of a set of control parameters, the Jarzynski equality reads\c{jarzynski1997nonequilibrium,jarzynskia2008nonequilibrium}
\beq
\label{jarz}
\av{\e^{-\beta W}}=\e^{-\beta \Delta F}.
\eeq
Here $W$ is work; the angle brackets denote an average over many independent dynamical trajectories resulting from the protocol; and $\Delta F$ is the free-energy difference associated with the initial and final values of the control parameters. However, \bb{because \eqq{jarz}} involves the average of an exponential, the number of trajectories required to estimate $\Delta F$ \bb{using it} grows exponentially with the variance of work fluctuations\c{jarzynski2006rare,hummer2010free,rohwer2015convergence} (see \s{var}). 

\begin{figure*} 
     \centering
     \includegraphics[width=\linewidth]{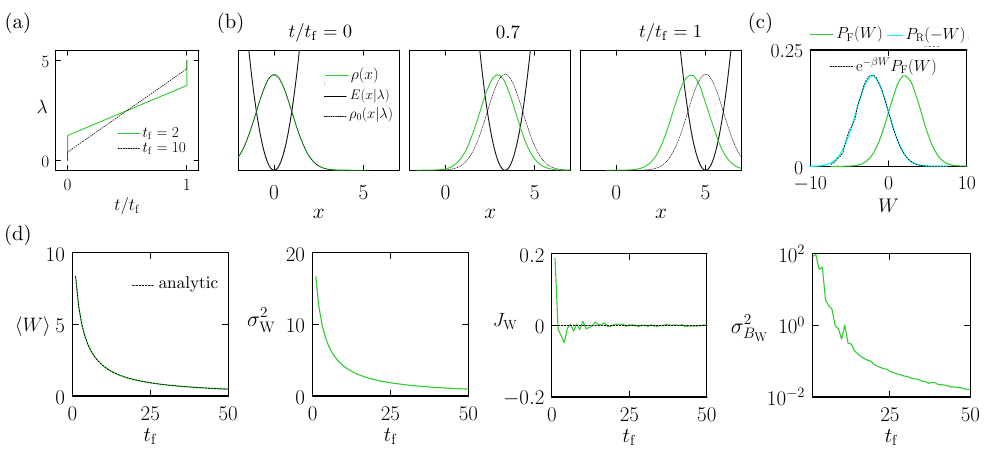} 
     \caption{Model of an overdamped colloidal particle in an optical trap, at constant temperature $\beta^{-1}=1$, translated according to the work-minimizing protocol of~\cc{schmiedl2007optimal}. (a) Protocol $\lambda^\star(t)$ for two values of trajectory length $\tf$. (b) Particle-position distribution, potential, and associated Boltzmann distribution, at three times $t$, for trajectories of length $\tf=10$. (c) Work statistics for this protocol and its time reverse satisfy the fluctuation relation \eq{fluc}. (d) Mean work $\av{W}$ (compared with the exact result $W^\star$\c{schmiedl2007optimal}, shown as a dashed line); work variance $\sigma^2_W$; Jarzynski free-energy estimator $J_W$; and variance $\sigma^2_{{\rm B}_W}$ of the block average of the exponential of $J_W$ (note the vertical log scale). Averages and distributions are calculated over $10^6$ trajectories.}
     \label{figa}
  \end{figure*}
  
This problem motivates the search for nonequilibrium protocols that perform desired transformations while minimizing work or other path-extensive quantities\c{schmiedl2007optimal,vaikuntanathan2008escorted,davie2014applicability,blaber2021steps,engel2022optimal,blaber2020skewed,shenfeld2009minimizing}~\footnote{Minimizing work does not necessarily minimize the fluctuations of work, but there is usually a strong correlation between these things: see e.g. panel (d) of \f{figa}.}. \bb{But while} protocols of this nature can involve varying temperature -- such as the protocol that reverses the magnetization of the Ising model with least dissipation\c{{rotskoff2015optimal,gingrich2016near}} -- \eqq{jarz} applies only at fixed temperature. 

In this paper we consider a variant of \eq{jarz} that applies to protocols whose temperature can vary with time. Such variants have been derived within the framework of Hamiltonian dynamics\c{williams2008nonequilibrium,chelli2009nonequilibrium,jarzynski1999microscopic} or Markovian dynamics specific to the Ising model\c{chatelain2007temperature}. We follow \cc{crooks1998nonequilibrium} and consider a general Markovian dynamics satisfying detailed balance. If we consider the protocol to involve a set of time-varying control parameters {\em and} a time-varying reciprocal temperature $\beta(t)$, with the latter starting and ending at a value $\beta$~\footnote{$\beta$ with no time argument or subscript denotes the fixed reciprocal temperature at the start and end of the trajectory: we want to estimate the value $\beta \Delta F$ appearing in \eq{jarz} by allowing the system at {\em intermediate} times to have a value $\beta(t)$ that may be different to the end-point value $\beta$.}, then \eq{jarz} is replaced by
\beq
\label{jmod}
\av{\e^{-\Omega}}=\e^{-\beta \Delta F}.
\eeq
Here $\Omega \equiv \beta W+\beta Q - \Sigma$, where $Q$ is the heat exchanged with the bath and $-\Sigma$ is the path entropy produced by the trajectory (the instantaneous change of heat divided by the instantaneous temperature, summed over the trajectory). The quantity $\Omega-\beta \Delta F$ is the total entropy production~\footnote{\eqq{jmod} can be considered a special case -- one where the temperatures at the trajectory endpoints are equal -- of a varying-temperature version of the entropy-production fluctuation theorem\c{evans2002fluctuation,seifert2012stochastic}}.

 The angle brackets in \eq{jmod} denote an average over nonequilibrium trajectories that start in equilibrium at temperature $\beta^{-1}$, that finish at the same temperature (not necessarily in equilibrium), and that otherwise involve an arbitrary change of temperature and other control parameters~\footnote{Very rapid temperature variation may drive the thermal bath out of equilibrium, in which case temperature is not a well-defined quantity\c{brey1990generalized}; the derivation assumes that the thermal bath remains in equilibrium.}. \bb{Similar results have been derived for Hamiltonian dynamics: \eqq{jmod} is equivalent to Eq. (9) of~\cc{jarzynski1999microscopic} (verified experimentally in \cc{rademacher2022nonequilibrium}) upon applying the first law of thermodynamics to that result and setting the start- and end temperatures equal.}

Under the same conditions, the fluctuation relation 
\beq
\label{fmod}
P_{\rm F}(\Omega) \hspace{0.3pt} \e^{-\Omega}=  \e^{-\beta \Delta F} P_{\rm R}(-\Omega)
\eeq
holds. Here $P_{\rm F}(\Omega)$ denotes the probability distribution of $\Omega$ under the protocol, and $P_{\rm R}(-\Omega)$ the distribution of $-\Omega$ under the time-reversed protocol. For a fixed-temperature trajectory we have $\beta Q  = \Sigma$ and so $\Omega=\beta W$, and \eq{jmod} and \eq{fmod} reduce to the Jarzynski equality \eq{jarz} and the Crooks fluctuation relation\c{crooks1999entropy}
\beq
\label{fluc}
P_{\rm F}(W) \hspace{0.3pt} \e^{-\beta W}=  \e^{-\beta \Delta F} P_{\rm R}(-W),
\eeq
respectively.  \bb{\eqq{fmod} can be obtained from the generic expression (1.12) of Ref.\c{crooks1999excursions} by imposing that we start in equilibrium, interact with a single bath with a time-varying temperature, and start and end at the same temperature. Related expressions for deterministic dynamics are given in Refs.\c{chelli2007generalization,chelli2007numerical}.} However, the specific forms \eq{jmod} and \eq{fmod} allow the extraction of free-energy differences at temperature $\beta^{-1}$, as do the original Jarzynski and Crooks relations, but with the potential for increased statistical accuracy.

In \s{sketch} we sketch the derivation of \eq{jmod} and \eq{fmod}. Details of the derivation are given in Appendix~\ref{gen}, which follows \cc{crooks1998nonequilibrium} with minor notational changes~\footnote{We also show that the Jarzynski equality becomes the staged Zwanzig formula for free-energy perturbation if the trajectory remains in equilibrium, and becomes the formula for thermodynamic integration if, in addition, the control parameters change in infinitesimal increments; related limiting forms were derived in~\cc{jarzynski1997nonequilibrium} within the framework of Hamiltonian dynamics.}. In \s{constant} we introduce a simulation model of a colloidal particle in an optical trap, and use it to show, in \s{simple}, that the fluctuation relations \eq{jmod} and \eq{fmod} hold for varying-temperature protocols. Notably, such protocols can give rise to smaller fluctuations of $\Omega$ and better convergence properties of \eq{jmod} than do fixed-temperature protocols for $W$ and \eq{jarz}, allowing more accurate extraction of free-energy differences. In \s{varying} we show that this difference is particularly pronounced in the presence of a phase transition, where the ability to vary temperature allows us to choose protocols that lead to much lower dissipation than the best fixed-temperature protocols. We conclude in \s{conclusions}.
  
\section{Sketch of derivation} 
\label{sketch}

    \begin{figure*} 
     \centering
     \includegraphics[width=\linewidth]{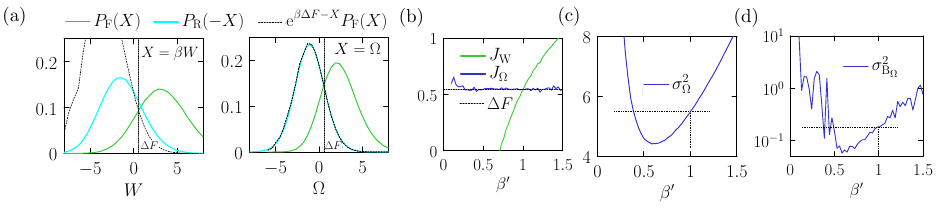} 
     \caption{Trap-translation protocols run at varying temperature $\beta(t)$, parameterized by $\beta'$ ($\beta'=1$ corresponds to constant temperature). (a) Probability distributions of $W$ and $\Omega$ for the protocol with $\beta'= 0.21$.  The work fluctuation relation \eq{fluc} does not hold (left panel), while the varying-temperature fluctuation relation \eq{fmod} does (right panel). (b) The logarithm of the estimator of \eq{jmod}, $J_\Omega$, approximates $\Delta F$, while the logarithm of the estimator of \eq{jarz}, $J_W$, does not. Fluctuations of (c) $\Omega$ and of (d) the block average of the estimator can be smaller for varying-temperature protocols ($\beta' \neq 1$) than for constant-temperature protocols ($\beta'=1$).}
     \label{figb}
  \end{figure*}

Consider a system at instantaneous temperature $\beta^{-1}(t)$. The system's microscopic coordinates are given by a vector $\xx$ and its energy function is $E(\xx|\llambda)$, where $\llambda$ is a vector of control parameters. We define the {\em protocol} as the deterministic time evolution of $\llambda(t)$ and $\beta(t)$. Starting in microstate $\xx(0)=\xx_0$ with the parameter values $\llambda(0)=\llambda_0$ and $\beta(0)=\beta_0$, a dynamical trajectory $\omega$ of the system consists of a series of alternating changes of the protocol, $\beta_0,\llambda_0 \to \beta_1,\llambda_1 \to \cdots \to \beta_N,\llambda_N$, and coordinates, $\xx_0 \to \xx_1 \to \cdots \to \xx_N$. If $P[0\to N]$ is the probability of generating the trajectory $\omega$, and $P[0\ot N]$ the probability of generating its reverse under the time-reversed protocol, then, for \bb{a} Markovian stochastic dynamics satisfying detailed balance, we have $P[0\to N]/P[0\ot N]=\e^{-\Sigma_\omega}$, where
\beq
\Sigma_\omega \equiv \sum_{i=0}^{N-1} \beta_{i+1} \left[E(\xx_{i+1}|\llambda_{i+1})-E(\xx_i|\llambda_{i+1}) \right]
\eeq
is (minus) the path entropy produced by $\omega$. \bb{This assumption requires the bath degrees of freedom to remain in equilibrium as temperature is varied, setting an upper limit on how rapidly this can be done\c{brey1990generalized}}. If we assume that $\omega$ and its reverse start in thermal equilibrium with respective control-parameter values $\beta_0,\llambda_0$ and $\beta_N,\llambda_N$, where $\beta_0=\beta_N \equiv \beta$, then the path-probability ratio can be written
\beq
\label{ratio}
P_0[0 \to N] \e^{-\beta W_\omega-\beta Q_\omega+\Sigma_\omega}  
= \e^{-\beta \Delta F} P_0[0 \ot N] .
\eeq
Here $\Delta F$ is the Helmholtz free energy difference at temperature $\beta^{-1}$ corresponding to the change $\llambda_0 \to \llambda_N$, and
\beq
W_\omega= \sum_{i=0}^{N-1} [E(\xx_i|\llambda_{i+1})-E(\xx_i|\llambda_i)]
 \eeq
and
 \beq
 \label{heat}
 Q_\omega = \sum_{i=0}^{N-1} \left[E(\xx_{i+1}|\llambda_{i+1})-E(\xx_i|\llambda_{i+1}) \right]
 \eeq
 are the work done and heat exchanged with the bath \bb{(the energy transferred to the system from the bath)} within $\omega$.
 
 Summing \eq{ratio} over all trajectories $\omega$ gives \eqq{jmod} (there we have dropped the path label subscript $\omega$ on $\Omega$). Multiplying both sizes of \eq{ratio} by $\delta{\left(\Omega_\omega-\Omega\right)}$ and summing over trajectories $\omega$ \bb{(once we have symmetrized the dynamics with respect to the order of state- and protocol changes)} gives \eqq{fmod}. 
 
 For details of this derivation, see Appendix~\ref{gen}.

\section{Constant-temperature protocols} 
\label{constant}

 Consider a model of an overdamped colloidal particle in an optical trap\c{schmiedl2007optimal}. The particle has position $x$ and the system has energy function $E(x|\lambda)=k(t)\left(x-\lambda(t) \right)^2/2$, where $\lambda$ specifies the trap center. The particle undergoes the Langevin dynamics
\beq
\label{langevin}
\dot{x}=-\partial_x E(x|\lambda) + \xi(t),
\eeq
which satisfies detailed balance with respect to the system's energy function\c{crooks1999excursions,dellago1998efficient}. The noise $\xi$ satisfies $\av{\xi(t)}=0$ and $\av{\xi(t) \xi(t')} = 2 \beta(t)^{-1} \delta(t-t')$. Initially we set $k(t)=1$ for all $t$.

We simulate \eq{langevin} using the forward Euler discretization with timestep $\Delta t=10^{-3}$. Starting in equilibrium at temperature $\beta^{-1}=1$, with the trap center at $\lambda(0)=\lambda_0=0$, we consider the fixed-temperature protocol that moves the trap center to a final position $\lambda(\tf) = \lambda_{\rm f}=5$, in finite time $t_{\rm f}$, and that minimizes the work averaged over many realizations of the process. This protocol has the form $\lambda^\star(t)=\lambda_{\rm f} (t+1)/(t_{\rm f}+2)$, for $0<t<t_{\rm f}$, with jump discontinuities at the start $(t=0)$ and end $(t=t_{\rm f})$, and produces mean work $W^\star=\lambda_{\rm f}^2/(t_{\rm f}+2)$\c{schmiedl2007optimal}. Note that $\Delta F=0$ for this protocol, because the energy function is translated but otherwise unchanged.

In \f{figa} we verify that work distributions produced by the protocol $\lambda^\star$ obey the standard relations \eq{jarz} and \eq{fluc}. In panel (a) we show the protocol $\lambda^\star$ for two trajectory lengths $\tf$. In panel (b) we show for $\tf=10$ the time-resolved distribution of particle positions $\rho(x)$ under this protocol, together with the energy function $E(x|\lambda)$ and the associated Boltzmann distribution $\rho_0(x|\lambda)$. Here and subsequently we calculate distributions and averages over $10^6$ independent trajectories. In panel (c) we verify that the work distribution produced by this protocol and its time reverse satisfy the work-fluctuation relation \eq{fluc}. 
\begin{figure*} 
     \centering
     \includegraphics[width=\linewidth]{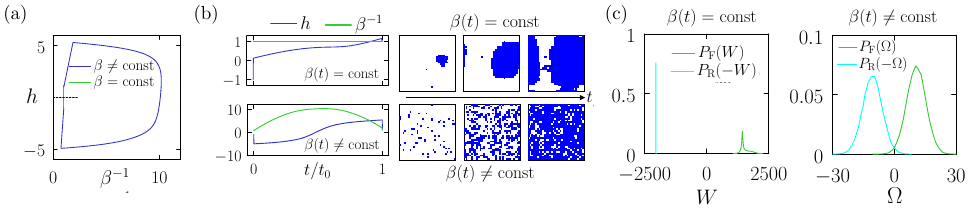} 
     \caption{Ising model magnetization reversal using dissipation-minimizing fixed- and varying-temperature protocols. Protocols were determined by genetic algorithm applied to a neural network\c{whitelam2023demon}. (a) Learned protocols in parameteric form. The dotted line is the first-order phase transition line. (b) Protocols as functions of time, together with typical time-ordered snapshots. (c) Fixed-temperature protocols (for which $\Omega = \beta W$) generate large amounts of dissipation, preventing application of the standard relations \eq{jarz} and \eq{fluc} (left panel). Varying-temperature protocols produce much less dissipation, allowing application of \eq{jmod} and \eq{fmod} (right panel).}
     \label{figc}
  \end{figure*}
  
In panel (d) we compare protocols $\lambda^\star$ carried out for a range of trajectory lengths $\tf$. Shown are the mean work $\av{W}$; the variance $\sigma^2_W$ of the work distribution; the Jarzynski free-energy estimator 
\beq
J_W =-\beta^{-1} \ln (N_{\rm traj}^{-1} \sum_{i=1}^{N_{\rm traj}} \e^{-\beta W_i}),
\eeq
where $i$ labels trajectories and $N_{\rm traj}=10^6$; and the variance $\sigma^2_{{\rm B}_W}$ of the block average ${\rm B}_W = N_{\rm block}^{-1} \sum_{i=1}^{N_{\rm block}} \e^{-\beta W_i}$, where $N_{\rm block}=100$. The mean work satisfies $\av{W}=W^\star$, as expected. The estimator $J_W$ returns $\Delta F=0$ for large values of $\tf$, but for small values of $\tf$ is imprecise; the fluctuation $\sigma^2_W$ is the source of this imprecision, and $\sigma^2_{{\rm B}_W}$ is one measure of its size.
  
\section{Simple varying-temperature protocols}
\label{simple}

 In \f{figb} we again consider the protocol $\lambda=\lambda^\star(t)$, for $\tf=10$, but now $\beta$ is varied in a piecewise-linear way, $\beta(t) = 1 + 2 (\beta' - 1) (t/\tf)$ for $t/\tf<1/2$ and $\beta(t)= \beta' + 2 (1 - \beta') (t/\tf- 1/2)$ for $t/\tf \geq 1/2$ (so that $\beta(0)=\beta(\tf)=1$). We consider a range of $\beta'$ either side of 1. We also consider a time-varying spring constant, $k(t) = (1-t/\tf)k +(t/\tf)k'$, which results in a free-energy difference $\Delta F =\frac{1}{2} \ln (k'/k)$: we choose $k'=3k=3$, giving $\Delta F \approx 0.55$.  In panel (a) we show that, as expected, the work fluctuation relation \eq{fluc} is not obeyed for a varying-temperature protocol (here $\beta'= 0.21$), but the varying-temperature fluctuation relation \eq{fmod} is. In panel (b) we show that, as expected, the standard Jarzynski equality no longer applies: the green line is the free-energy estimator $J_W$, which for $\beta' \neq 1$ does not equal $\Delta F$. By contrast, the estimator 
 \beq
 J_\Omega =-\beta^{-1} \ln (N_{\rm traj}^{-1} \sum_{i=1}^{N_{\rm traj}} \e^{-\Omega_i})
 \eeq 
 provides a noisy estimate of $\Delta F$, indicating that \eq{jmod} is obeyed. 
 
In panels (c) and (d)  we show, as a function of $\beta'$, the variance $\sigma^2_W$ of $\Omega$ and the variance $\sigma^2_{{\rm B}_\Omega}$ of the block average ${\rm B}_\Omega = N_{\rm block}^{-1} \sum_{i=1}^{N_{\rm block}} \e^{-\Omega_i}$, with $N_{\rm block}=100$. The minimum of the latter occurs around $\beta'=0.2$, and is less than half that at $\beta'=1$ (where $\Omega=\beta W$ and the relations \eq{jmod} and \eq{fmod} reduce to \eq{jarz} and \eq{fluc}). While fluctuations tend to increase with temperature, the combination $\Omega \equiv \beta W+\beta Q-\Sigma$ and its fluctuations can be smaller than $W$ and its fluctuations. These effects compete, and for some range of $\beta' \neq 1$ the fluctuations of $\Omega$ are smaller than those of $W$ at $\beta'=1$, leading to better convergence of \eq{jmod} than \eq{jarz}.

\section{Varying-temperature protocols in the presence of a phase transition}
\label{varying}

This difference in convergence properties is small, because the physics of the trap model does not change substantially with temperature. But when a system's physics does change with temperature, the difference between the convergence properties of the fixed- and varying-temperature relations can be significant. 
 
 In \f{figc} we consider \bb{magnetization reversal in the the 2D ferromagnetic Ising model, a simple model of nanomagnetic bit copying and erasure\c{rotskoff2015optimal,gingrich2016near}. We consider a $32 \times 32$ lattice, with fixed coupling $J=1$, and simulate the model} using Glauber dynamics for $\tf=10^3$ Monte Carlo sweeps. Protocols start and end at parameter values $\beta(0)=\beta(\tf)=1$ and $h(0)=-h(\tf)=-1$, giving $\Delta F=0$, in which case $\Omega$ is equal to the total entropy production. We determine time-dependent protocols $\beta(t)$ and $h(t)$ for $0 < t < \tf$ by expressing these quantities using a deep neural network and training that network by genetic algorithm to minimize $\av{\Omega}$ measured over $10^4$ independent trajectories\c{whitelam2023demon}. Protocols learned in this way effect magnetization reversal. We consider one set of learning simulations with the constraint that $\beta(t)$ remain constant, and another in which $\beta(t)$ is allowed to vary.
 
 In panel (a) we show in parametric form the protocols learned in this way. The varying-temperature protocol avoids the critical point and the first-order phase transition line\c{rotskoff2015optimal,gingrich2016near}, while the fixed-temperature protocol is constrained to cross it. In panel (b) we show protocols as functions of time, together with time-ordered snapshots taken from typical trajectories. The fixed-temperature protocol effects nucleation and growth, accompanied by large values of dissipation: $\av{\Omega}= \av{\beta  W} \approx 1550$. By contrast, the varying-temperature protocol is accompanied by much smaller dissipation, with $\av{\Omega} \approx 11$. 
 
 As a result, distributions of work for the fixed-temperature protocol and its time reverse are well-separated, while distributions of $\Omega$ for the varying-temperature protocol and its time reverse cross at a value that approximates $\Delta F=0$; see panel (c). Using $10^5$ trajectories, the free-energy estimator $J_\Omega$ yields a value 0.46, with block-average variance $\sigma^2_{{\rm B}_\Omega} \approx 13,700$. Thus \eq{fmod} and \eq{jmod} provide an accurate if imprecise measure of $\Delta F$. By contrast, if we are constrained to fixed temperature in order to apply the standard relations \eq{jarz} and \eq{fluc}, measuring $\Delta F$ is not a realistic proposition. Fixed-temperature trajectories hundreds of times longer would be required to achieve convergence properties comparable to \eq{jmod} and \eq{fmod} using varying-temperature trajectories~\footnote{We could change $J$ at fixed $\beta$ in order to mimic temperature variation, but our model study is carried out to represent an experiment in which we cannot change the microscopic parameters of a material, while we can use temperature as a control parameter to take us across a phase boundary.}.

\section{Conclusions}
\label{conclusions} 

We have shown, within the framework of Markovian stochastic dynamics satisfying detailed balance, that the relations \eq{jmod} and \eq{fmod} replace the Jarzynski equality \eq{jarz} and Crooks work fluctuation relation \eq{fluc}, for trajectories influenced by a time-varying temperature that starts and ends at a value $\beta^{-1}$. We have used simulation models to show that free-energy differences can be calculated more accurately using the varying-temperature relations than the fixed-temperature ones, particularly when varying temperature gives us the freedom to avoid the large dissipation associated with a first-order phase transition. To measure the quantity $\Omega=\beta W + \beta Q - \Sigma$ that appears in \eq{jmod} and \eq{fmod} we must be able to measure the time-resolved heat flow, as well as total heat and work. For many experiments this will be technically demanding, but it is possible in principle. 
 
\section{Acknowledgments} 

I thank Corneel Casert and David Sivak for comments on the paper, and Fr\'ed\'eric van Wijland for bringing~\cc{brey1990generalized} to my attention. Code for the trap model can be found here\c{trap_github}. Code for doing neuroevolutionary learning of Ising model protocols can be found here\c{demon_github}, which accompanies~\cc{whitelam2023demon} (for this paper we neglect the shear term, Eq. (7), used in that work). This work was performed at the Molecular Foundry at Lawrence Berkeley National Laboratory, supported by the Office of Basic Energy Sciences of the U.S. Department of Energy under Contract No. DE-AC02--05CH11231.


%

 \appendix
 
 \renewcommand{\thefigure}{A\arabic{figure}}
\renewcommand{\thesection}{A\arabic{section}}
\renewcommand{\theequation}{A\arabic{equation}}

\setcounter{equation}{0}
\setcounter{section}{0}
\setcounter{figure}{0}

\section{Work fluctuations and the Jarzynski equality}
\label{var}

 To illustrate in a simple way the convergence problems of an exponential average, assume that the distribution $P(W)$ of nonequilibrium work values is Gaussian with mean $\bar{W}$ and variance $\sigma^2$, $P(W) \propto \e^{-(W-\bar{W})^2/(2 \sigma^2)}$. The average on the left-hand size of \eq{jarz} can then be written 
 \beq
 \av{\e^{-\beta W}} \sim \int {\rm d} W \e^{-(W-W_{\rm a})^2/(2 \sigma^2)},
 \eeq 
 which is dominated by contributions from the atypical work value $W=W_{\rm a} \equiv \bar{W} - \beta \sigma^2$. The probability of realizing this work value is $P(W_{\rm a})\propto \e^{-\beta^2 \sigma^2/2}$, which requires a characteristic number of trajectories $\sim 1/P(W_{\rm a}) \propto \e^{\beta^2 \sigma^2/2}$. This quantity grows exponentially with the variance $\sigma^2$ of work fluctuations.

Work distributions are in general not Gaussian, but it is usually the case that the larger the fluctuations of $W$, the more trajectories are required to calculate \eq{jarz}. 

 \section{Constant-temperature protocols}
 \label{gen}
 
\subsection{Markovian stochastic dynamics satisfying detailed balance}
\label{dev}
 
 \begin{figure*} 
   \centering
 \includegraphics[width=0.75\linewidth]{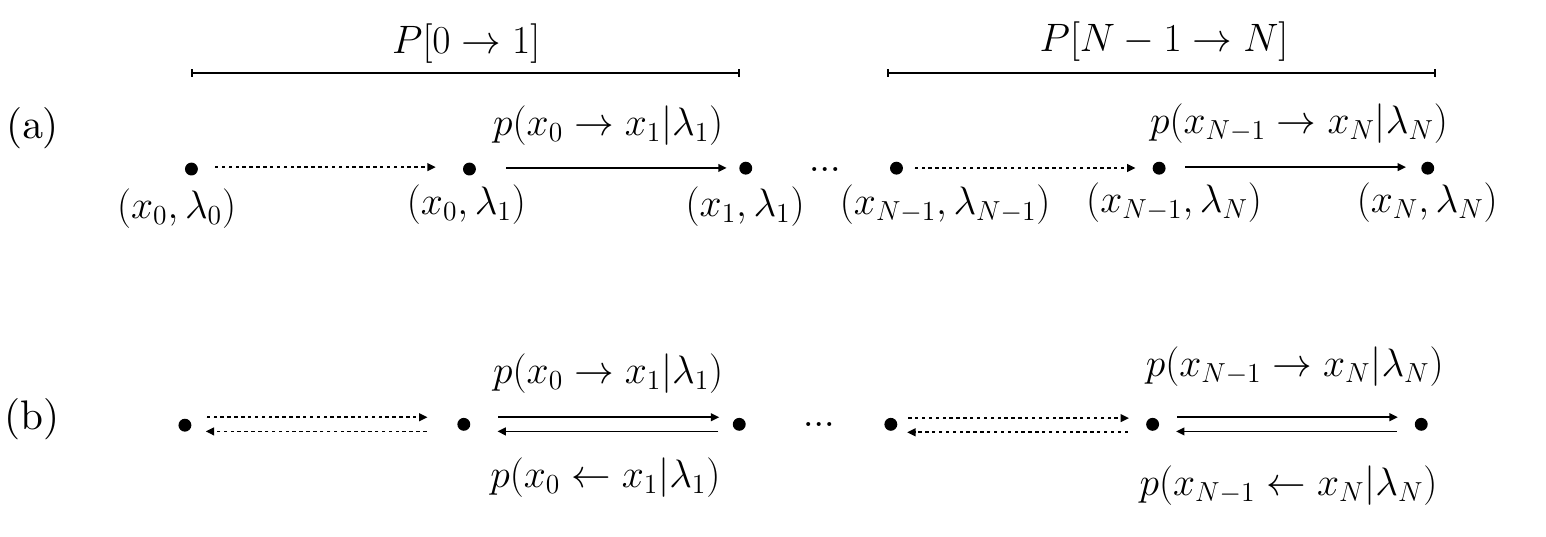} 
   \caption{(a) Forward trajectory, and (b) forward-reverse trajectory pair.}
   \label{fig1}
\end{figure*}

In this supplement we derive the expressions \eq{jmod} and \eq{fmod} used in the main text, following~\cc{crooks1998nonequilibrium} with minor notational changes. We consider a stochastic, Markovian dynamics that satisfies detailed balance at temperature $\beta^{-1}$ with respect to the energy function $E(\xx|\llambda)$. Here $\xx$ is the vector of microscopic coordinates of the system, and $\llambda$ is a vector of control parameters. As shown in \f{fig1}(a), a dynamical trajectory of the system involves $N$ deterministic changes of the control-parameter vector $\llambda$, according to $\llambda_0 \to \llambda_1 \to \cdots \to \llambda_N$. The system coordinates $\xx$ evolve stochastically as $\xx_0 \to \xx_1 \to \cdots \to \xx_N$. We consider these changes to occur in an alternating fashion: $\llambda$ changes along dotted arrows, with work done or expended, and $\xx$ changes along solid arrows, with heat exchanged with the thermal bath. \bb{Here we follow~\cc{crooks1998nonequilibrium} in choosing a specific ordering for state- and protocol changes; in \s{sec_fluc} we shall symmetrize the trajectory with respect to the order of these changes.}

The probability with which the trajectory of \f{fig1}(a) occurs is $P[0 \to N]$, where
\bea
\label{prod}
P[a \to b] &=& \prod_{i=a}^{b-1}P[i \to i+1] \nonumber \\ 
&=& \prod_{i=a}^{b-1} p(\xx_i \to \xx_{i+1}|\llambda_{i+1}).
\eea
Here $p(\xx_i \to \xx_{i+1}|\llambda_{i+1})$ is the probability of moving from microstate $\xx_i$ to microstate $\xx_{i+1}$, given that the control-parameter vector is $\llambda_{i+1}$. The product structure of \eq{prod} follows from the Markovian property of the dynamics. 

As a convenient device, \cc{crooks1998nonequilibrium} introduces the notion of a \bb{time-reversed} trajectory, shown as the lower set of arrows in \f{fig1}(b), in which $\llambda$ and $\xx$ evolve in the reverse order to the forward trajectory. The ratio of path probabilities of forward and reverse trajectories is
\bea
\label{path}
\frac{P[0 \to N]}{P[0 \ot N]}&=&\prod_{i=0}^{N-1} \frac{p(\xx_i \to \xx_{i+1}|\llambda_{i+1})}{p(\xx_i \ot \xx_{i+1}|\llambda_{i+1})} \\
\label{path2}
&=&\prod_{i=0}^{N-1} \e^{-\beta(E(\xx_{i+1}|\llambda_{i+1})-E(\xx_i|\llambda_{i+1}))} \\
\label{microrev}
&=&\e^{-\beta Q_{0 \to N}}.
\eea
Here $p(\xx_i \ot \xx_{i+1}|\llambda_{i+1})$ is the probability of moving from microstate $\xx_{i+1}$ to microstate $\xx_i$, given that the control-parameter vector is $\llambda_{i+1}$, and
 \beq
 \label{heat}
 Q_{0 \to N} = \sum_{i=0}^{N-1} \left[E(\xx_{i+1}|\llambda_{i+1})-E(\xx_i|\llambda_{i+1}) \right]
 \eeq
 is the heat exchanged with the bath \bb{(the energy transferred to the system from the bath)} along the forward trajectory. In the main text we use the subscript $\omega$ to indicate a trajectory-dependent quantity. Here we use the more detailed notation $0 \to N$, so that we can in describe portions of the trajectory using the notation $i \to j$. The passage from \eq{path} to \eq{path2} follows from the fact that the dynamics satisfies detailed balance. \eqq{microrev} is a statement of microscopic reversibility: it is guaranteed if the dynamics satisfies detailed balance, but also holds if the dynamics satisfied global balance and not detailed balance.
 
Given a control-parameter vector $\llambda_i$, the likelihood of observing microstate $\xx_j$ in thermal equilibrium is 
\beq
\rho(\xx_j|\llambda_i) = \e^{\beta (F_\beta(\llambda_i)-E(\xx_j|\llambda_i))},
\eeq
where
\beq
F_\beta(\llambda_i)=-\beta^{-1} \ln \sum_\xx \e^{-\beta E(\xx|\llambda_i)}
\eeq
is the Helmholtz free energy of the system under control-parameter vector $\llambda_i$ (we shall drop subscript labels $\beta$ on $F$ unless considering free energies calculated at different temperatures). If we assume that forward and reverse trajectories each start in thermal equilibrium under respective control-parameter values $\llambda_0$ and $\llambda_N$, then the path-probability ratio \eq{path} becomes
\beq
\label{path3}
\frac{\rho(\xx_0|\llambda_0)P[0 \to N]}{P[0 \ot N]\rho(\xx_N|\llambda_N)}=\e^{-\beta (\Delta F_{0 \to N} -\Delta E_{0 \to N}+Q_{0 \to N})},
\eeq
 where $\Delta F_{0 \to N} \equiv F(\llambda_N) - F(\llambda_0)$ and $\Delta E_{0 \to N} \equiv E(\xx_N|\llambda_N) - E(\xx_0|\llambda_0)$. Using the first law of thermodynamics,
 \beq
 Q_{0 \to N}+W_{0 \to N}=\Delta E_{0 \to N},
 \eeq
 where 
 \beq
 \label{work}
W_{0 \to N}= \sum_{i=0}^{N-1} [E(\xx_i|\llambda_{i+1})-E(\xx_i|\llambda_i)]
 \eeq
 is the work done along the forward trajectory, \eq{path3} can be written
 \beq
\label{path4}
\frac{\rho(\xx_0|\llambda_0)P[0 \to N]}{P[0 \ot N]\rho(\xx_N|\llambda_N)}=\e^{-\beta (\Delta F_{0 \to N} -W_{0 \to N})}.
\eeq
It will be convenient to write \eq{path4} as
 \bea
\label{path5}
\rho(\xx_0|\llambda_0)&P&[0 \to N] \e^{-\beta W_{0 \to N}}\nonumber \\&=&\e^{-\beta \Delta F_{0 \to N}} P[0 \ot N]\rho(\xx_N|\llambda_N).
\eea

 \subsection{Jarzynski equality}
 \label{je}
 
Summing \eq{path5} over all possible trajectories $\{ \xx\}$ gives
\bea
\label{int}
&\,&\sum_{\{ \xx\}} \rho(\xx_0|\llambda_0)P[0 \to N] \e^{-\beta W_{0 \to N}} \nonumber \\&=& \e^{-\beta \Delta F_{0 \to N}} \sum_{\{ \xx\} } P[0 \ot N]\rho(\xx_N|\llambda_N).
\eea 
The sum on the right-hand side of \eq{int} is unity, by normalization of probabilities, and we can write what remains as
\beq
\label{jarzprime}
\av{ \e^{-\beta W_{0 \to N}}}_{\llambda_0 \to \llambda_N} = \e^{-\beta \Delta F_{0 \to N}}.
\eeq
The angle brackets in \eq{jarzprime} denotes an average over trajectories under the forward protocol, starting from thermal equilibrium at the control-parameter vector $\llambda_0$. This completes the proof of the Jarzynski equality given in \cc{crooks1998nonequilibrium}, for a Markovian, stochastic dynamics satisfying detailed balance. Only the starting point of the forward trajectory is in thermal equilibrium; no subsequent points on the trajectory, including the final one, need be in equilibrium. The \bb{time-reversed} trajectory is introduced as a device to ensure a convenient cancellation of terms in the derivation of \eq{jarzprime}, but no reverse trajectories need be considered for its calculation. 

In the remainder of \s{gen} we consider some special cases and limits of the Jarzynski equality.
 \begin{figure*} 
   \centering
 \includegraphics[width=0.85\linewidth]{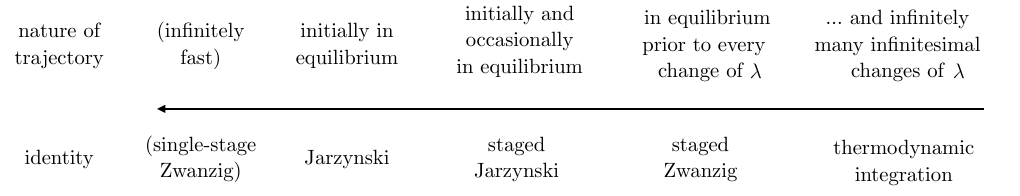} 
   \caption{Summary of the results of \s{gen}, relating the nature of the forward trajectory of \f{fig1}(a) to the identity that applies to it. The line points in the direction of increasing rate of transformation.}
   \label{fig2}
\end{figure*}

 \subsection{Staged Jarzynski equality}
 \label{sje}

Assume now that forward and reverse trajectories attain equilibrium when the control-parameter vector is $\llambda_i$, where $0<i<N$. To enforce this assumption we insert the factor $\rho(\xx_i|\llambda_i)$ in the numerator and denominator of \eq{path4}, giving
 \beq
\label{path6}
\frac{\rho(\xx_0|\llambda_0)P[0 \to i]\rho(\xx_i|\llambda_i)P[i \to N]}{P[0 \ot i]\rho(\xx_i|\llambda_i) P[i \ot N]\rho(\xx_N|\llambda_N)}.
\eeq
The value of \eq{path6} is again equal to $\e^{-\beta (\Delta F_{0 \to N} -W_{0 \to N})}$, but forward and reverse trajectories now possess a constraint not present in the previous case. This constraint could be enforced by having a time-varying protocol that pauses at the value $\llambda_i$ for as long as required to achieve equilibrium. However, we could also consider the expression \eq{path6} to refer to two sets of trajectory pairs that involve the portions $0 \to i$ and $i \to N$ of the original trajectory, respectively. These trajectory pairs start in equilibrium and possess no additional constraints. The form of \eq{path6} consistent with this assumption is
 \beq
\label{path7}
\frac{\rho(\xx_0|\llambda_0)P[0 \to i]}{P[0 \ot i]\rho(\xx_i|\llambda_i)}\cdot \frac{\rho(\xx_i|\llambda_i)P[i \to N]}{P[i \ot N]\rho(\xx_N|\llambda_N)},
\eeq
the two factors in \eq{path7} corresponding to the two trajectory pairs. \bb{The first factor is}
\beq
\label{path7b}
\bb{\frac{\rho(\xx_0|\llambda_0)P[0 \to i]}{P[0 \ot N]\rho(\xx_i|\llambda_i)}=\e^{-\beta (\Delta F_{0 \to i}-W_{0 \to i})},}
\eeq
\bb{which can be rearranged and summed over ${\xx}$ to give}
\beq
\bb{\av{ \e^{-\beta W_{0 \to i}}}_{\llambda_0 \to \llambda_i} =\e^{-\beta \Delta F_{0 \to i}}.}
\eeq
\bb{As before, the angle brackets denote averages over all trajectories, starting in equilibrium under control-parameter vector $\llambda_0$.
The second factor in \eq{path7} gives}
\beq
\bb{\av{ \e^{-\beta W_{i \to N}}}_{\llambda_i \to \llambda_N} =\e^{-\beta \Delta F_{i \to N}}.}
\eeq
\bb{Because free energies are additive, i.e. $\Delta F_{0 \to i}+\Delta F_{i \to N}=\Delta F_{0 \to N}$, we can write}
\beq
\label{path8}
\av{ \e^{-\beta W_{0 \to i}}}_{\llambda_0 \to \llambda_i} \av{ \e^{-\beta W_{i \to N}}}_{\llambda_i \to \llambda_N} = \e^{-\beta \Delta F_{0 \to N}},
\eeq
\eqq{path8} is the statement that the Jarzynski equality \eq{jarzprime} can be evaluated by staging\c{pohorille2010good}, dividing a single trajectory (which starts in equilibrium but need not be in equilibrium subsequently) into shorter trajectories (each of which starts in equilibrium but need not be in equilibrium subsequently). This is obvious on physical grounds, given that the Jarzynski equality is a method for evaluating free-energy differences, and it applies to trajectories of arbitrary length. 

\subsection{Staged Zwanzig formula for free-energy perturbation}
\label{sz}

A special case arises when we insert, in the numerator and denominator of \eq{path4}, factors of $\rho(\xx_i|\llambda_i)$ for {\em all} values of $i$, so assuming that both trajectories attain equilibrium between each variation of the control-parameter vector $\llambda$ and the next. In this case \eq{path4} becomes
 \beq
 \label{zw1}
\prod_{i=0}^{N-1} \frac{\rho(\xx_i|\llambda_i)P[i \to i+1]}{P[i \ot i+1]\rho(\xx_{i+1}|\llambda_{i+1})}=\e^{-\beta \Delta F_{0 \to N}}.
\eeq 
Each factor on the left-hand side of \eq{zw1} is
\bea
\label{zw2}
\frac{\rho(\xx_i|\llambda_i)P[i \to i+1]}{P[i \ot i+1]\rho(\xx_{i+1}|\llambda_{i+1})}&=&\e^{-\beta (E(\xx_i|\llambda_i)-E(\xx_i|\llambda_{i+1}))} \nonumber \\
&\times& \e^{-\beta(F(\llambda_{i+1})-F(\llambda_i))}.
\eea
The right-hand side of \eq{zw2} depends on the energy at a single coordinate $\xx_i$ only, and so the notion of an explicit dynamics is absent. Rearranging \eq{zw2} and summing over trajectories $\{\xx\}$ gives
\bea
&\,&\sum_{\xx_i} \rho(\xx_i|\llambda_i)\e^{-\beta (E(\xx_i|\llambda_{i+1})-E(\xx_i|\llambda_i))} \sum_{\xx_{i+1}} P[i \to i+1]\nonumber \\
&=& \e^{-\beta (F(\llambda_{i+1})-F(\llambda_i))} 
\sum_{\xx_i,\xx_{i+1}} P[i \ot i+1]\rho(\xx_{i+1}|\llambda_{i+1}), \nonumber
\eea
which can be written 
\beq
\label{zw3}
\av{\e^{-\beta (E(\xx|\llambda_{i+1})-E(\xx|\llambda_i))}}_{\llambda_i}=\e^{-\beta \Delta F_{i \to i+1}}.
\eeq
Here the angle brackets $\av{(\cdot)}_{\llambda_i}$ denote an equilibrium average $\sum_\xx \rho(\xx|\llambda_i) (\cdot)$ under the control-parameter vector $\llambda_i$. \eqq{zw3} is the exponential of the Zwanzig formula for free-energy perturbation\c{zwanzig1954high}. Using \eq{zw3}, \eq{zw1} can be written
\beq
\label{pert}
\Delta F_{0 \to N} =-\beta^{-1} \sum_{i=0}^{N-1} \ln \av{\e^{-\beta (E(\xx|\llambda_{i+1})-E(\xx|\llambda_i))}}_{\llambda_i},
\eeq
which is a staged version of Zwanzig formula for $N$ changes of the control-parameter vector $\llambda$. To calculate \eq{pert}, it is natural to consider $N$ independent trajectories that all begin in equilibrium and consist of a single change of the control-parameter vector.

It was shown in \cc{jarzynski1997nonequilibrium} that the Jarzynski equality reduces to the Zwanzig formula 
\beq
\label{pert2}
\Delta F_{0 \to N} =-\beta^{-1} \ln \av{\e^{-\beta (E(\xx|\llambda_{0})-E(\xx|\llambda_N))}}_{\llambda_0}
\eeq
for a single, instantaneous change of the control-parameter vector from $\llambda_0 \to \llambda_N$. \eqq{pert} is the staged variant of this expression: \eq{pert} applies if the transformation is done in well-separated stages, with the system coming to equilibrium after each control-parameter change. 

\subsection{Thermodynamic integration}
\label{tint}

If we further assume that the trajectory involves only small changes of the control-parameter vector, $\llambda_{i+1} = \llambda_i + \delta \llambda$, such that 
\beq
E(\xx|\llambda_{i+1}) \approx E(\xx|\llambda_i)+ \delta \llambda \cdot \frac{\partial E(\xx|\llambda)}{\partial \llambda}|_{\llambda=\llambda_i},
\eeq
then \eq{pert} becomes
\bea
\Delta F_{0 \to N} &\approx& -\beta^{-1} \sum_{i=0}^{N-1} \ln \left( 1- \beta \delta \llambda \cdot \av{\frac{\partial E(\xx|\llambda)}{\partial \llambda}}_{\llambda_i} \right) \nonumber \\
&\approx& \sum_{i=0}^{N-1} \delta \llambda \cdot \av{\frac{\partial E(\xx|\llambda)}{\partial \llambda}}_{\llambda_i}.
\eea
In the limit of a large number $N \to \infty$ of vanishingly small changes $\delta \llambda \to 0$ we can write the above as
\beq
\label{ti}
\Delta F_{0 \to N} = \int_{\llambda_0}^{\llambda_N} {\rm d}\llambda \cdot \av{\frac{\partial E(\xx|\llambda)}{\partial \llambda}}_\llambda,
\eeq
which is the formula for thermodynamic integration\c{kirkwood1935statistical,frenkel2001understanding}. In \cc{jarzynski1997nonequilibrium} it was shown, within the framework of Hamiltonian dynamics, that thermodynamic integration is recovered from the Jarzynski equality in the limit of an infinitely slow transformation.

\subsection{Summary of this section}
 \begin{figure*} 
   \centering
 \includegraphics[width=0.75\linewidth]{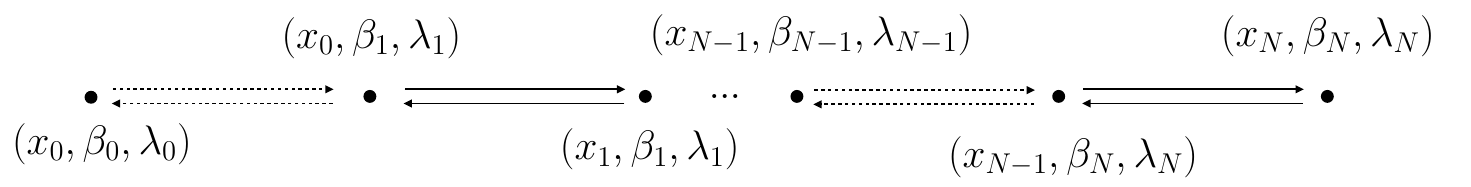} 
   \caption{Modification of the protocol of \f{fig1} to permit a time-varying temperature $\beta_i^{-1}$.}
   \label{fig3}
\end{figure*}

\f{fig2} summarizes the results of \s{gen}. In \cc{crooks1998nonequilibrium} it was shown that consideration of forward and reverse trajectory-pairs permits a simple proof of the Jarzynski equality\c{jarzynski1997nonequilibrium}, \eqq{jarzprime}, for trajectories of a Markovian stochastic dynamics that satisfies detailed balance, provided that these trajectories begin in thermal equilibrium with respect to the initial value of the control-parameter vector.  We have summarized this proof in \s{dev} and \s{je}. Using the same framework we have shown in \s{sje} that if trajectories also attain thermal equilibrium with other values of the control-parameter vector then we recover the (physically obvious) statement that the Jarzynski equality can be evaluated in a staged way. In \s{sz} we show that if trajectories attain thermal equilibrium after {\em all} changes of the control-parameter vector then the same considerations yield a staged version of the Zwanzig formula\c{zwanzig1954high} for free-energy perturbation, \eqq{pert}. (The single-stage Zwanzig formula is recovered in the limit of a single, instantaneous change of $\llambda$\c{jarzynski1997nonequilibrium}.) In \s{tint} we show that if, in addition, the control-parameter vector is changed in a large number of infinitesimally small steps, we recover the formula for thermodynamic integration, \eqq{ti}. (The same formula is recovered within the framework of Hamiltonian dynamics in the limit of an infinitely slow transformation\c{jarzynski1997nonequilibrium}.)

\section{Varying-temperature protocols}
\label{sec_vary}

In this section we modify the proof of \s{gen} to allow for a time-varying reciprocal temperature $\beta$, such that $\beta_0 \to \beta_1 \to \cdots \to \beta_N$ along a trajectory. We change $\beta$ in step with the control-parameter vector $\llambda$, as shown in \f{fig3}, and so where $\llambda=\llambda_i$ we have $\beta = \beta_i$. We note that the microscopic energies $E(\xx_i|\llambda_i)$ could in principle be temperature dependent. 

Under this new protocol, \eqq{microrev} reads
\bea
\label{microrev_mod}
\frac{P[0 \to N]}{P[0 \ot N]}&=&\prod_{i=0}^{N-1} \frac{p(\xx_i \to \xx_{i+1}|\llambda_{i+1})}{p(\xx_i \ot \xx_{i+1}|\llambda_{i+1})} \nonumber \\
\label{path2b}
&=&\prod_{i=0}^{N-1} \e^{-\beta_{i+1}(E(\xx_{i+1}|\llambda_{i+1})-E(\xx_i|\llambda_{i+1}))} \nonumber \\
\label{newpath}
&=&\e^{-\Sigma_{0 \to N}},
\eea
where
\beq
 \label{entprod}
\Sigma_{0 \to N} \equiv \sum_{i=0}^{N-1} \beta_{i+1} \left[E(\xx_{i+1}|\llambda_{i+1})-E(\xx_i|\llambda_{i+1}) \right]
\eeq
is (minus) the path entropy \bb{within} the forward trajectory (i.e. neglecting the endpoint distributions). \eqq{entprod} is an explicit realization of the path term appearing in the generic form $\Delta S_{\rm baths}$ given in Eq. (1.18) of \cc{crooks1999excursions}. \bb{These expressions assume the thermal bath to remain in equilibrium at all times; see \cc{brey1990generalized} for a discussion of a finite-size bath that relaxes at finite rate.}

For a varying-temperature protocol, \eqq{path3} becomes
\bea
\label{int_1}
\frac{\rho(\xx_0|\llambda_0)P[0 \to N]}{P[0 \ot N]\rho(\xx_N|\llambda_N)}&=&\e^{\beta_0 F_{\beta_0}(\llambda_0)-\beta_N F_{\beta_N}(\llambda_N)} \\
&\times& \e^{\beta_N E(\xx_N|\llambda_N)-\beta_0 E(\xx_0|\llambda_0)-\Sigma_{0 \to N}} \nonumber.
\eea
Because our goal is to evaluate free-energy differences at fixed temperature $\beta^{-1}$, we now specify that the temperatures at the start and end of the trajectory are equal, $\beta_0=\beta_N \equiv \beta$ (we allow general $\beta_i>0$ for $0<i<N$). In this case the free-energy difference appearing in the first line of \eq{int_1} is again $-\beta \Delta F_{0 \to N}$, and the difference of internal energies given on the second line of \eq{int_1} becomes $\beta \Delta E_{0 \to N}$. This can be eliminated in favor of the path-dependent combination $\beta(W_{0\to N}+Q_{0\to N})$: $\beta$ times the sum of \eq{heat} and \eq{work} is equal to $\beta \Delta E_{0 \to N}$, whether or not the time-dependent temperature along the path is equal to the temperature at the trajectory endpoints.

Thus, for a varying-temperature trajectory that starts and ends at temperature $\beta^{-1}$, \eqq{int_1} can be written
\bea
\frac{\rho(\xx_0|\llambda_0)P[0 \to N]}{P[0 \ot N]\rho(\xx_N|\llambda_N)}&=&\e^{-\beta \Delta F_{0 \to N} +\beta \Delta E_{0 \to N}-\Sigma_{0 \to N}} \\
&=&\e^{-\beta \Delta F_{0 \to N} +\beta (W_{0 \to N}+Q_{0 \to N})-\Sigma_{0 \to N}}, \nonumber
\eea
giving
\bea
\label{tj1}
\rho(\xx_0|\llambda_0)P[0 \to N] \e^{-\beta W_{0 \to N}-\beta Q_{0 \to N}+\Sigma_{0 \to N}}  \nonumber \\
= P[0 \ot N]\rho(\xx_N|\llambda_N) \e^{-\beta \Delta F_{0 \to N}}.
\eea
Summing \eq{tj1} over all trajectories $\{\xx\}$ gives 
\beq
\label{jarz_mod}
\av{ \e^{-\Omega_{0 \to N}}}_{\llambda_0 \to \llambda_N} = \e^{-\beta \Delta F_{0 \to N}},
\eeq
where 
\beq
\label{sigma}
\Omega_{0 \to N} \equiv \beta W_{0 \to N}+\beta Q_{0 \to N}-\Sigma_{0 \to N}.
\eeq
The angle brackets in \eq{jarz_mod} denote an average over nonequilibrium trajectories of fixed but arbitrary length that start in equilibrium with reciprocal temperature $\beta_0=\beta$ and control-parameter vector $\llambda_0$, end with reciprocal temperature $\beta_N=\beta$ and control-parameter vector $\llambda_N$, and otherwise involve an arbitrary time-dependent variation of $\beta_i$ and $\llambda_i$.

\eqq{jarz_mod}, which is \eqq{jmod} of the main text, is a variant of the Jarzynski equality \eq{jarzprime} that is valid for a varying-temperature protocol with equal start- and end temperatures.

The Jensen inequality applied to \eq{jarz_mod} yields $\av{\Omega_{0 \to N}} \geq \beta \Delta F$, the statement that the total entropy production is nonnegative. This is an expression of the second law of thermodynamics, \bb{analogous} to the statement $\av{W} \geq \Delta F$ that results from Jensen's inequality applied to the Jarzynski equality.

\section{Fluctuation relations}
\label{sec_fluc}
 
 \subsection{Fixed temperature}
 
In this section we consider the fluctuation relations that correspond to the expressions \eq{jarzprime} and \eq{jarz_mod}. Thus far, time-reversed trajectories have been considered as a device to derive the Jarzynski equality and related identities, but evaluation of those identities requires only the generation of trajectories using the forward protocol. In this subsection we will consider expressions (again derived using time-reversed trajectories as a convenient device) that refer explicitly to the ensemble of trajectories generated using the normal dynamics into which the time-reversed protocol has been inserted.

So far we have followed \cc{crooks1998nonequilibrium} in considering a stochastic dynamics in which, as shown in \f{fig1}, protocol changes and state changes alternate, but occur in fixed order. The forward trajectory shown in that figure starts with a protocol change, and so the time-reversed trajectory starts with a state change. In order to derive the Jarzynski identity (or the relation \eq{jarz_mod} in the varying-temperature case), we need only that the time-reversed dynamics is normalized, which it is. We do not need to generate that dynamics explicitly. However, the derivation of \eq{fluc}, the Crooks relation\c{crooks1999entropy} (or of \eq{fmod} in the varying-temperature case), requires that we associate the time-reversed trajectory with a trajectory generated by the normal dynamics using the time-reversed protocol. To do so, we need to consider a dynamics that is symmetrized with respect to the order of state- and protocol changes.

To see this, note that we can multiply \eq{path5} by $\delta{\left(W_{0 \to N}-W\right)}$ and sum over all possible trajectories $\{ \xx\}$. The result is
 \bea
\label{int2}
&\,&\sum_{\{ \xx\}} \delta{\left(W_{0 \to N}-W\right)} \rho(\xx_0|\llambda_0)P[0 \to N] \e^{-\beta W_{0 \to N}} \\&=& \e^{-\beta \Delta F_{0 \to N}} \sum_{\{ \xx\} } \delta{\left(W_{0 \to N}-W\right)} P[0 \ot N]\rho(\xx_N|\llambda_N). \nonumber
\eea 
We can write this as
\beq
\label{fluc1}
P_{\rm F}(W) \, \e^{-\beta W} =  \e^{-\beta \Delta F_{0 \to N}} P_{\rm R}(-W),
\eeq
where $P_{\rm F}(W)$ is the probability of observing work value $W_{0 \to N}=W$ for a trajectory generated by the forward protocol $\llambda(t)$. $P_{\rm R}(-W)$ is the probability of observing work value $W_{N \to 0} = -W$ for the ensemble of time-reversed trajectories (note that $W_{0 \to N}=-W_{N \to 0}$). The quantity $P_{\rm R}(-W)$ is not automatically the same thing as the probability distribution of $W$ for the normal dynamics into which the time-reversed protocol $\llambda(\tf-t)$ has been inserted, because the latter dynamics always begins with a protocol change and so could never generate the lower (reverse) trajectory shown in \f{fig1}(b), which begins with a state change. 

In order to sidestep this issue we can symmetrize the dynamics with respect to state- and protocol changes, defining it so that the first step of a trajectory is, with equal probability, a state change or a protocol change. We can then interpret the reverse trajectory drawn in \f{fig1} as a trajectory generated by the normal dynamics into which the time-reversed protocol has been inserted. Given the trajectories drawn in \f{fig1}, the right-hand side of \eqq{prod} picks up a factor of $1/2$ (because with that probability we choose the ordering shown), but so too does the expression $P[0 \leftarrow N]$ for the reverse trajectory, and so \eqq{path} remains unchanged. As before, we proceed to \eqq{path5}. We multiply this equation by $\delta{\left(W_{0 \to N}-W\right)}$ and sum over $\{ \xx\}$, giving
\beq
\label{fluc1_mod}
P_{\rm F}^{(1)}(W) \, \e^{-\beta W} =  \e^{-\beta \Delta F_{0 \to N}} P_{\rm R}^{(2)}(-W).
\eeq
Here the probability distributions F and R are over trajectories generated by the dynamics under the forward $\llambda(t)$ and time-reversed protocols $\llambda(\tf-t)$, respectively, while the superscripts 1 and 2 indicate the subset of dynamical trajectories in which a protocol change or a state change is proposed first. 

There is then a diagram similar to that shown in \eq{fig1}(b) but with the order of state changes and protocol changes swapped. This gives rise to the same series of equations as before, with the bookkeeping change that heat and work are now defined as
\beq
 Q_{0 \to N} = \sum_{i=0}^{N-1} \left[E(\xx_{i+1}|\llambda_{i})-E(\xx_i|\llambda_{i}) \right]
 \eeq
and
 \beq
W_{0 \to N}= \sum_{i=0}^{N-1} [E(\xx_{i+1}|\llambda_{i+1})-E(\xx_{i+1}|\llambda_i)].
 \eeq
Once again, $Q_{0 \to N} +W_{0 \to N} = \Delta E_{0 \to N} \equiv E(\xx_N|\llambda_N) - E(\xx_0|\llambda_0)$. We then obtain a version of \eqq{fluc1_mod} with superscript labels reversed,
\beq
\label{fluc1_mod2}
P_{\rm F}^{(2)}(W) \, \e^{-\beta W} =  \e^{-\beta \Delta F_{0 \to N}} P_{\rm R}^{(1)}(-W).
\eeq
Adding \eq{fluc1_mod} and  \eq{fluc1_mod} gives 
\beq
\label{fluc1mod}
P_{\rm F}(W) \, \e^{-\beta W} =  \e^{-\beta \Delta F_{0 \to N}} P_{\rm R}(-W),
\eeq
which is \eqq{fluc} of the main text, once we note that $P_{\rm F}(W)=(P_{\rm F}^{(1)}(W)+P_{\rm F}^{(2)}(W))/2$ and $P_{\rm R}(-W)=(P_{\rm R}^{(1)}(-W)+P_{\rm R}^{(2)}(-W))/2$. Now, however, we can interpret the symbol $P_{\rm R}(-W)$ in \eq{fluc1mod} as the probability distribution of $-W$ under the normal dynamics using the time-reversed protocol $\llambda(\tf-t)$, which can be straightforwardly calculated in simulations.

The difference between symmetric and asymmetric forms of the dynamics is likely negligible for all but the shortest trajectories, but we note that it is more natural to associate the time-reversed dynamics with the normal dynamics under the time-reversed protocol if the latter can start with a state change {\em or} a protocol change.

 \subsection{Varying temperature}

 Similar considerations apply to the derivation of the fluctuation relations corresponding to \eq{jarz_mod}. We consider a dynamics that starts with either a state change or a protocol change, with equal probability. \eqq{microrev_mod} acquires factors of $1/2$ in its numerator and denominator, and so remains unchanged. We then end up with two versions of \eqq{tj1}, one for the ordering of state- and protocol changes shown in \f{fig3}, and one for its reverse. Multiplying each by $\delta{\left(W_{0 \to N}-W\right)} \delta{\left(Q_{0 \to N}-Q\right)} \delta{\left(\Sigma_{0 \to N}-\Sigma\right)}$, summing over $\{ \xx\}$, and adding the resulting equations gives the detailed fluctuation relation
\bea
\label{fluc2}
P_{\rm F}(W,&Q&,\Sigma) \, \e^{-\beta W-\beta Q+\Sigma} \nonumber \\
&=&  \e^{-\beta \Delta F_{0 \to N}} P_{\rm R}(-W,-Q,-\Sigma).
\eea
Here $P_{\rm F}(W,Q,\Sigma)$ is the joint probability of observing the values $(W_{0 \to N},Q_{0 \to N},\Sigma_{0 \to N})=(W,Q,\Sigma)$ under the forward protocol $(\beta(t),\llambda(t))$, while $P_{\rm R}(-W,-Q,-\Sigma)$ is is the joint probability of observing the values $(W_{0 \to N},Q_{0 \to N},\Sigma_{0 \to N})=(-W,-Q,-\Sigma)$ under the time-reversed protocol $(\beta(\tf-t),\llambda(\tf-t))$.

The simple fluctuation relation associated with \eq{jarz_mod} follows by multiplying both versions of \eq{tj1} by $\delta{\left(\Omega_{0 \to N}-\Omega\right)}$, summing over $\{ \xx\}$, and adding the resulting equations, and is
\beq
\label{fluc3}
P_{\rm F}(\Omega) \, \e^{-\Omega} =  \e^{-\beta \Delta F_{0 \to N}} P_{\rm R}(-\Omega),
\eeq
which is \eqq{fmod} of the main text. Here $P_{\rm F}(\Omega)$ is the probability of observing the value $\Omega_{0 \to N}=\Omega$ under the forward protocol $(\beta(t),\llambda(t))$, and $P_{\rm R}(-\Omega)$ is the probability of observing the value $\Omega_{N \to 0} = -\Omega$ under the time-reversed protocol $(\beta(\tf-t),\llambda(\tf-t))$; note that $\Omega_{0 \to N}=-\Omega_{N \to 0}$.  Integrating \eq{fluc3} over $\Omega$ gives \eq{jarz_mod}. 

For a constant-temperature protocol, $\beta Q = \Sigma$ and $\Omega=\beta W$, and both \eq{fluc2} and \eq{fluc3} reduce to \eq{fluc1mod}.

Finally, note that $\Omega-\beta \Delta F_{0 \to N}$ is the total entropy production, and \eq{fluc3} can be considered a special case (one where the temperatures at the trajectory endpoints are equal) of a varying-temperature version of the entropy-production fluctuation theorem\c{evans2002fluctuation}.


\begin{thebibliography}{41}%
\makeatletter
\providecommand \@ifxundefined [1]{%
 \@ifx{#1\undefined}
}%
\providecommand \@ifnum [1]{%
 \ifnum #1\expandafter \@firstoftwo
 \else \expandafter \@secondoftwo
 \fi
}%
\providecommand \@ifx [1]{%
 \ifx #1\expandafter \@firstoftwo
 \else \expandafter \@secondoftwo
 \fi
}%
\providecommand \natexlab [1]{#1}%
\providecommand \enquote  [1]{``#1''}%
\providecommand \bibnamefont  [1]{#1}%
\providecommand \bibfnamefont [1]{#1}%
\providecommand \citenamefont [1]{#1}%
\providecommand \href@noop [0]{\@secondoftwo}%
\providecommand \href [0]{\begingroup \@sanitize@url \@href}%
\providecommand \@href[1]{\@@startlink{#1}\@@href}%
\providecommand \@@href[1]{\endgroup#1\@@endlink}%
\providecommand \@sanitize@url [0]{\catcode `\\12\catcode `\$12\catcode
  `\&12\catcode `\#12\catcode `\^12\catcode `\_12\catcode `\%12\relax}%
\providecommand \@@startlink[1]{}%
\providecommand \@@endlink[0]{}%
\providecommand \url  [0]{\begingroup\@sanitize@url \@url }%
\providecommand \@url [1]{\endgroup\@href {#1}{\urlprefix }}%
\providecommand \urlprefix  [0]{URL }%
\providecommand \Eprint [0]{\href }%
\providecommand \doibase [0]{http://dx.doi.org/}%
\providecommand \selectlanguage [0]{\@gobble}%
\providecommand \bibinfo  [0]{\@secondoftwo}%
\providecommand \bibfield  [0]{\@secondoftwo}%
\providecommand \translation [1]{[#1]}%
\providecommand \BibitemOpen [0]{}%
\providecommand \bibitemStop [0]{}%
\providecommand \bibitemNoStop [0]{.\EOS\space}%
\providecommand \EOS [0]{\spacefactor3000\relax}%
\providecommand \BibitemShut  [1]{\csname bibitem#1\endcsname}%
\let\auto@bib@innerbib\@empty
\bibitem [{\citenamefont {Jarzynski}(1997)}]{jarzynski1997nonequilibrium}%
  \BibitemOpen
  \bibfield  {author} {\bibinfo {author} {\bibfnamefont {Christopher}\
  \bibnamefont {Jarzynski}},\ }\bibfield  {title} {\enquote {\bibinfo {title}
  {Nonequilibrium equality for free energy differences},}\ }\href@noop {}
  {\bibfield  {journal} {\bibinfo  {journal} {Physical Review Letters}\
  }\textbf {\bibinfo {volume} {78}},\ \bibinfo {pages} {2690} (\bibinfo {year}
  {1997})}\BibitemShut {NoStop}%
\bibitem [{\citenamefont {Jarzynski}(2008)}]{jarzynskia2008nonequilibrium}%
  \BibitemOpen
  \bibfield  {author} {\bibinfo {author} {\bibfnamefont {C}~\bibnamefont
  {Jarzynski}},\ }\bibfield  {title} {\enquote {\bibinfo {title}
  {Nonequilibrium work relations: foundations and applications},}\ }\href@noop
  {} {\bibfield  {journal} {\bibinfo  {journal} {The European Physical Journal
  B}\ }\textbf {\bibinfo {volume} {64}},\ \bibinfo {pages} {331--340} (\bibinfo
  {year} {2008})}\BibitemShut {NoStop}%
\bibitem [{\citenamefont {Jarzynski}(2006)}]{jarzynski2006rare}%
  \BibitemOpen
  \bibfield  {author} {\bibinfo {author} {\bibfnamefont {Christopher}\
  \bibnamefont {Jarzynski}},\ }\bibfield  {title} {\enquote {\bibinfo {title}
  {Rare events and the convergence of exponentially averaged work values},}\
  }\href@noop {} {\bibfield  {journal} {\bibinfo  {journal} {Physical Review
  E}\ }\textbf {\bibinfo {volume} {73}},\ \bibinfo {pages} {046105} (\bibinfo
  {year} {2006})}\BibitemShut {NoStop}%
\bibitem [{\citenamefont {Hummer}\ and\ \citenamefont
  {Szabo}(2010)}]{hummer2010free}%
  \BibitemOpen
  \bibfield  {author} {\bibinfo {author} {\bibfnamefont {Gerhard}\ \bibnamefont
  {Hummer}}\ and\ \bibinfo {author} {\bibfnamefont {Attila}\ \bibnamefont
  {Szabo}},\ }\bibfield  {title} {\enquote {\bibinfo {title} {Free energy
  profiles from single-molecule pulling experiments},}\ }\href@noop {}
  {\bibfield  {journal} {\bibinfo  {journal} {Proceedings of the National
  Academy of Sciences}\ }\textbf {\bibinfo {volume} {107}},\ \bibinfo {pages}
  {21441--21446} (\bibinfo {year} {2010})}\BibitemShut {NoStop}%
\bibitem [{\citenamefont {Rohwer}\ \emph {et~al.}(2015)\citenamefont {Rohwer},
  \citenamefont {Angeletti},\ and\ \citenamefont
  {Touchette}}]{rohwer2015convergence}%
  \BibitemOpen
  \bibfield  {author} {\bibinfo {author} {\bibfnamefont {Christian~M}\
  \bibnamefont {Rohwer}}, \bibinfo {author} {\bibfnamefont {Florian}\
  \bibnamefont {Angeletti}}, \ and\ \bibinfo {author} {\bibfnamefont {Hugo}\
  \bibnamefont {Touchette}},\ }\bibfield  {title} {\enquote {\bibinfo {title}
  {Convergence of large-deviation estimators},}\ }\href@noop {} {\bibfield
  {journal} {\bibinfo  {journal} {Physical Review E}\ }\textbf {\bibinfo
  {volume} {92}},\ \bibinfo {pages} {052104} (\bibinfo {year}
  {2015})}\BibitemShut {NoStop}%
\bibitem [{\citenamefont {Schmiedl}\ and\ \citenamefont
  {Seifert}(2007)}]{schmiedl2007optimal}%
  \BibitemOpen
  \bibfield  {author} {\bibinfo {author} {\bibfnamefont {Tim}\ \bibnamefont
  {Schmiedl}}\ and\ \bibinfo {author} {\bibfnamefont {Udo}\ \bibnamefont
  {Seifert}},\ }\bibfield  {title} {\enquote {\bibinfo {title} {Optimal
  finite-time processes in stochastic thermodynamics},}\ }\href@noop {}
  {\bibfield  {journal} {\bibinfo  {journal} {Physical Review Letters}\
  }\textbf {\bibinfo {volume} {98}},\ \bibinfo {pages} {108301} (\bibinfo
  {year} {2007})}\BibitemShut {NoStop}%
\bibitem [{\citenamefont {Vaikuntanathan}\ and\ \citenamefont
  {Jarzynski}(2008)}]{vaikuntanathan2008escorted}%
  \BibitemOpen
  \bibfield  {author} {\bibinfo {author} {\bibfnamefont {Suriyanarayanan}\
  \bibnamefont {Vaikuntanathan}}\ and\ \bibinfo {author} {\bibfnamefont
  {Christopher}\ \bibnamefont {Jarzynski}},\ }\bibfield  {title} {\enquote
  {\bibinfo {title} {Escorted free energy simulations: Improving convergence by
  reducing dissipation},}\ }\href@noop {} {\bibfield  {journal} {\bibinfo
  {journal} {Physical Review Letters}\ }\textbf {\bibinfo {volume} {100}},\
  \bibinfo {pages} {190601} (\bibinfo {year} {2008})}\BibitemShut {NoStop}%
\bibitem [{\citenamefont {Davie}\ \emph {et~al.}(2014)\citenamefont {Davie},
  \citenamefont {Jepps}, \citenamefont {Rondoni}, \citenamefont {Reid},\ and\
  \citenamefont {Searles}}]{davie2014applicability}%
  \BibitemOpen
  \bibfield  {author} {\bibinfo {author} {\bibfnamefont {Stuart~J}\
  \bibnamefont {Davie}}, \bibinfo {author} {\bibfnamefont {Owen~G}\
  \bibnamefont {Jepps}}, \bibinfo {author} {\bibfnamefont {Lamberto}\
  \bibnamefont {Rondoni}}, \bibinfo {author} {\bibfnamefont {James~C}\
  \bibnamefont {Reid}}, \ and\ \bibinfo {author} {\bibfnamefont {Debra~J}\
  \bibnamefont {Searles}},\ }\bibfield  {title} {\enquote {\bibinfo {title}
  {Applicability of optimal protocols and the {J}arzynski equality},}\
  }\href@noop {} {\bibfield  {journal} {\bibinfo  {journal} {Physica Scripta}\
  }\textbf {\bibinfo {volume} {89}},\ \bibinfo {pages} {048002} (\bibinfo
  {year} {2014})}\BibitemShut {NoStop}%
\bibitem [{\citenamefont {Blaber}\ \emph {et~al.}(2021)\citenamefont {Blaber},
  \citenamefont {Louwerse},\ and\ \citenamefont {Sivak}}]{blaber2021steps}%
  \BibitemOpen
  \bibfield  {author} {\bibinfo {author} {\bibfnamefont {Steven}\ \bibnamefont
  {Blaber}}, \bibinfo {author} {\bibfnamefont {Miranda~D}\ \bibnamefont
  {Louwerse}}, \ and\ \bibinfo {author} {\bibfnamefont {David~A}\ \bibnamefont
  {Sivak}},\ }\bibfield  {title} {\enquote {\bibinfo {title} {Steps minimize
  dissipation in rapidly driven stochastic systems},}\ }\href@noop {}
  {\bibfield  {journal} {\bibinfo  {journal} {Physical Review E}\ }\textbf
  {\bibinfo {volume} {104}},\ \bibinfo {pages} {L022101} (\bibinfo {year}
  {2021})}\BibitemShut {NoStop}%
\bibitem [{\citenamefont {Engel}\ \emph {et~al.}(2023)\citenamefont {Engel},
  \citenamefont {Smith},\ and\ \citenamefont {Brenner}}]{engel2022optimal}%
  \BibitemOpen
  \bibfield  {author} {\bibinfo {author} {\bibfnamefont {Megan~C}\ \bibnamefont
  {Engel}}, \bibinfo {author} {\bibfnamefont {Jamie~A}\ \bibnamefont {Smith}},
  \ and\ \bibinfo {author} {\bibfnamefont {Michael~P}\ \bibnamefont
  {Brenner}},\ }\bibfield  {title} {\enquote {\bibinfo {title} {Optimal control
  of nonequilibrium systems through automatic differentiation},}\ }\href@noop
  {} {\bibfield  {journal} {\bibinfo  {journal} {Physical Review X}\ }\textbf
  {\bibinfo {volume} {13}},\ \bibinfo {pages} {041032} (\bibinfo {year}
  {2023})}\BibitemShut {NoStop}%
\bibitem [{\citenamefont {Blaber}\ and\ \citenamefont
  {Sivak}(2020)}]{blaber2020skewed}%
  \BibitemOpen
  \bibfield  {author} {\bibinfo {author} {\bibfnamefont {Steven}\ \bibnamefont
  {Blaber}}\ and\ \bibinfo {author} {\bibfnamefont {David~A}\ \bibnamefont
  {Sivak}},\ }\bibfield  {title} {\enquote {\bibinfo {title} {Skewed
  thermodynamic geometry and optimal free energy estimation},}\ }\href@noop {}
  {\bibfield  {journal} {\bibinfo  {journal} {The Journal of Chemical Physics}\
  }\textbf {\bibinfo {volume} {153}} (\bibinfo {year} {2020})}\BibitemShut
  {NoStop}%
\bibitem [{\citenamefont {Shenfeld}\ \emph {et~al.}(2009)\citenamefont
  {Shenfeld}, \citenamefont {Xu}, \citenamefont {Eastwood}, \citenamefont
  {Dror},\ and\ \citenamefont {Shaw}}]{shenfeld2009minimizing}%
  \BibitemOpen
  \bibfield  {author} {\bibinfo {author} {\bibfnamefont {Daniel~K}\
  \bibnamefont {Shenfeld}}, \bibinfo {author} {\bibfnamefont {Huafeng}\
  \bibnamefont {Xu}}, \bibinfo {author} {\bibfnamefont {Michael~P}\
  \bibnamefont {Eastwood}}, \bibinfo {author} {\bibfnamefont {Ron~O}\
  \bibnamefont {Dror}}, \ and\ \bibinfo {author} {\bibfnamefont {David~E}\
  \bibnamefont {Shaw}},\ }\bibfield  {title} {\enquote {\bibinfo {title}
  {Minimizing thermodynamic length to select intermediate states for
  free-energy calculations and replica-exchange simulations},}\ }\href@noop {}
  {\bibfield  {journal} {\bibinfo  {journal} {Physical Review E}\ }\textbf
  {\bibinfo {volume} {80}},\ \bibinfo {pages} {046705} (\bibinfo {year}
  {2009})}\BibitemShut {NoStop}%
\bibitem [{Note1()}]{Note1}%
  \BibitemOpen
  \bibinfo {note} {Minimizing work does not necessarily minimize the
  fluctuations of work, but there is usually a strong correlation between these
  things: see e.g. panel (d) of Fig.~\ref {figa}.}\BibitemShut {Stop}%
\bibitem [{\citenamefont {Rotskoff}\ and\ \citenamefont
  {Crooks}(2015)}]{rotskoff2015optimal}%
  \BibitemOpen
  \bibfield  {author} {\bibinfo {author} {\bibfnamefont {Grant~M}\ \bibnamefont
  {Rotskoff}}\ and\ \bibinfo {author} {\bibfnamefont {Gavin~E}\ \bibnamefont
  {Crooks}},\ }\bibfield  {title} {\enquote {\bibinfo {title} {Optimal control
  in nonequilibrium systems: Dynamic {R}iemannian geometry of the {I}sing
  model},}\ }\href@noop {} {\bibfield  {journal} {\bibinfo  {journal} {Physical
  Review E}\ }\textbf {\bibinfo {volume} {92}},\ \bibinfo {pages} {060102}
  (\bibinfo {year} {2015})}\BibitemShut {NoStop}%
\bibitem [{\citenamefont {Gingrich}\ \emph {et~al.}(2016)\citenamefont
  {Gingrich}, \citenamefont {Rotskoff}, \citenamefont {Crooks},\ and\
  \citenamefont {Geissler}}]{gingrich2016near}%
  \BibitemOpen
  \bibfield  {author} {\bibinfo {author} {\bibfnamefont {Todd~R}\ \bibnamefont
  {Gingrich}}, \bibinfo {author} {\bibfnamefont {Grant~M}\ \bibnamefont
  {Rotskoff}}, \bibinfo {author} {\bibfnamefont {Gavin~E}\ \bibnamefont
  {Crooks}}, \ and\ \bibinfo {author} {\bibfnamefont {Phillip~L}\ \bibnamefont
  {Geissler}},\ }\bibfield  {title} {\enquote {\bibinfo {title} {Near-optimal
  protocols in complex nonequilibrium transformations},}\ }\href@noop {}
  {\bibfield  {journal} {\bibinfo  {journal} {Proceedings of the National
  Academy of Sciences}\ }\textbf {\bibinfo {volume} {113}},\ \bibinfo {pages}
  {10263--10268} (\bibinfo {year} {2016})}\BibitemShut {NoStop}%
\bibitem [{\citenamefont {Williams}\ \emph {et~al.}(2008)\citenamefont
  {Williams}, \citenamefont {Searles},\ and\ \citenamefont
  {Evans}}]{williams2008nonequilibrium}%
  \BibitemOpen
  \bibfield  {author} {\bibinfo {author} {\bibfnamefont {Stephen~R}\
  \bibnamefont {Williams}}, \bibinfo {author} {\bibfnamefont {Debra~J}\
  \bibnamefont {Searles}}, \ and\ \bibinfo {author} {\bibfnamefont {Denis~J}\
  \bibnamefont {Evans}},\ }\bibfield  {title} {\enquote {\bibinfo {title}
  {Nonequilibrium free-energy relations for thermal changes},}\ }\href@noop {}
  {\bibfield  {journal} {\bibinfo  {journal} {Physical Review Letters}\
  }\textbf {\bibinfo {volume} {100}},\ \bibinfo {pages} {250601} (\bibinfo
  {year} {2008})}\BibitemShut {NoStop}%
\bibitem [{\citenamefont {Chelli}(2009)}]{chelli2009nonequilibrium}%
  \BibitemOpen
  \bibfield  {author} {\bibinfo {author} {\bibfnamefont {Riccardo}\
  \bibnamefont {Chelli}},\ }\bibfield  {title} {\enquote {\bibinfo {title}
  {Nonequilibrium work relations for systems subject to mechanical and thermal
  changes},}\ }\href@noop {} {\bibfield  {journal} {\bibinfo  {journal} {The
  Journal of Chemical Physics}\ }\textbf {\bibinfo {volume} {130}} (\bibinfo
  {year} {2009})}\BibitemShut {NoStop}%
\bibitem [{\citenamefont {Jarzynski}(1999)}]{jarzynski1999microscopic}%
  \BibitemOpen
  \bibfield  {author} {\bibinfo {author} {\bibfnamefont {C}~\bibnamefont
  {Jarzynski}},\ }\bibfield  {title} {\enquote {\bibinfo {title} {Microscopic
  analysis of {C}lausius--{D}uhem processes},}\ }\href@noop {} {\bibfield
  {journal} {\bibinfo  {journal} {Journal of Statistical Physics}\ }\textbf
  {\bibinfo {volume} {96}},\ \bibinfo {pages} {415--427} (\bibinfo {year}
  {1999})}\BibitemShut {NoStop}%
\bibitem [{\citenamefont {Chatelain}(2007)}]{chatelain2007temperature}%
  \BibitemOpen
  \bibfield  {author} {\bibinfo {author} {\bibfnamefont {Christophe}\
  \bibnamefont {Chatelain}},\ }\bibfield  {title} {\enquote {\bibinfo {title}
  {A temperature-extended {J}arzynski relation: application to the numerical
  calculation of surface tension},}\ }\href@noop {} {\bibfield  {journal}
  {\bibinfo  {journal} {Journal of Statistical Mechanics: Theory and
  Experiment}\ }\textbf {\bibinfo {volume} {2007}},\ \bibinfo {pages} {P04011}
  (\bibinfo {year} {2007})}\BibitemShut {NoStop}%
\bibitem [{\citenamefont {Crooks}(1998)}]{crooks1998nonequilibrium}%
  \BibitemOpen
  \bibfield  {author} {\bibinfo {author} {\bibfnamefont {Gavin~E}\ \bibnamefont
  {Crooks}},\ }\bibfield  {title} {\enquote {\bibinfo {title} {Nonequilibrium
  measurements of free energy differences for microscopically reversible
  markovian systems},}\ }\href@noop {} {\bibfield  {journal} {\bibinfo
  {journal} {Journal of Statistical Physics}\ }\textbf {\bibinfo {volume}
  {90}},\ \bibinfo {pages} {1481--1487} (\bibinfo {year} {1998})}\BibitemShut
  {NoStop}%
\bibitem [{Note2()}]{Note2}%
  \BibitemOpen
  \bibinfo {note} {$\beta $ with no time argument or subscript denotes the
  fixed reciprocal temperature at the start and end of the trajectory: we want
  to estimate the value $\beta \Delta F$ appearing in (\ref {jarz}) by allowing
  the system at {\protect \em intermediate} times to have a value $\beta (t)$
  that may be different to the end-point value $\beta $.}\BibitemShut {Stop}%
\bibitem [{Note3()}]{Note3}%
  \BibitemOpen
  \bibinfo {note} {Eq.~(\ref {jmod}) can be considered a special case -- one
  where the temperatures at the trajectory endpoints are equal -- of a
  varying-temperature version of the entropy-production fluctuation
  theorem~\cite {evans2002fluctuation,seifert2012stochastic}}\BibitemShut
  {NoStop}%
\bibitem [{Note4()}]{Note4}%
  \BibitemOpen
  \bibinfo {note} {Very rapid temperature variation may drive the thermal bath
  out of equilibrium, in which case temperature is not a well-defined
  quantity~\cite {brey1990generalized}; the derivation assumes that the thermal
  bath remains in equilibrium.}\BibitemShut {Stop}%
\bibitem [{\citenamefont {Rademacher}\ \emph {et~al.}(2022)\citenamefont
  {Rademacher}, \citenamefont {Konopik}, \citenamefont {Debiossac},
  \citenamefont {Grass}, \citenamefont {Lutz},\ and\ \citenamefont
  {Kiesel}}]{rademacher2022nonequilibrium}%
  \BibitemOpen
  \bibfield  {author} {\bibinfo {author} {\bibfnamefont {Markus}\ \bibnamefont
  {Rademacher}}, \bibinfo {author} {\bibfnamefont {Michael}\ \bibnamefont
  {Konopik}}, \bibinfo {author} {\bibfnamefont {Maxime}\ \bibnamefont
  {Debiossac}}, \bibinfo {author} {\bibfnamefont {David}\ \bibnamefont
  {Grass}}, \bibinfo {author} {\bibfnamefont {Eric}\ \bibnamefont {Lutz}}, \
  and\ \bibinfo {author} {\bibfnamefont {Nikolai}\ \bibnamefont {Kiesel}},\
  }\bibfield  {title} {\enquote {\bibinfo {title} {Nonequilibrium control of
  thermal and mechanical changes in a levitated system},}\ }\href@noop {}
  {\bibfield  {journal} {\bibinfo  {journal} {Physical Review Letters}\
  }\textbf {\bibinfo {volume} {128}},\ \bibinfo {pages} {070601} (\bibinfo
  {year} {2022})}\BibitemShut {NoStop}%
\bibitem [{\citenamefont {Crooks}(1999{\natexlab{a}})}]{crooks1999entropy}%
  \BibitemOpen
  \bibfield  {author} {\bibinfo {author} {\bibfnamefont {Gavin~E}\ \bibnamefont
  {Crooks}},\ }\bibfield  {title} {\enquote {\bibinfo {title} {Entropy
  production fluctuation theorem and the nonequilibrium work relation for free
  energy differences},}\ }\href@noop {} {\bibfield  {journal} {\bibinfo
  {journal} {Physical Review E}\ }\textbf {\bibinfo {volume} {60}},\ \bibinfo
  {pages} {2721} (\bibinfo {year} {1999}{\natexlab{a}})}\BibitemShut {NoStop}%
\bibitem [{\citenamefont {Crooks}(1999{\natexlab{b}})}]{crooks1999excursions}%
  \BibitemOpen
  \bibfield  {author} {\bibinfo {author} {\bibfnamefont {Gavin~E.}\
  \bibnamefont {Crooks}},\ }\href@noop {} {\emph {\bibinfo {title} {Excursions
  in statistical dynamics}}}\ (\bibinfo  {publisher} {PhD Thesis, University of
  California, Berkeley},\ \bibinfo {year} {1999})\BibitemShut {NoStop}%
\bibitem [{\citenamefont {Chelli}\ \emph
  {et~al.}(2007{\natexlab{a}})\citenamefont {Chelli}, \citenamefont {Marsili},
  \citenamefont {Barducci},\ and\ \citenamefont
  {Procacci}}]{chelli2007generalization}%
  \BibitemOpen
  \bibfield  {author} {\bibinfo {author} {\bibfnamefont {Riccardo}\
  \bibnamefont {Chelli}}, \bibinfo {author} {\bibfnamefont {Simone}\
  \bibnamefont {Marsili}}, \bibinfo {author} {\bibfnamefont {Alessandro}\
  \bibnamefont {Barducci}}, \ and\ \bibinfo {author} {\bibfnamefont {Piero}\
  \bibnamefont {Procacci}},\ }\bibfield  {title} {\enquote {\bibinfo {title}
  {Generalization of the {J}arzynski and {C}rooks nonequilibrium work theorems
  in molecular dynamics simulations},}\ }\href@noop {} {\bibfield  {journal}
  {\bibinfo  {journal} {Physical Review E}\ }\textbf {\bibinfo {volume} {75}},\
  \bibinfo {pages} {050101} (\bibinfo {year} {2007}{\natexlab{a}})}\BibitemShut
  {NoStop}%
\bibitem [{\citenamefont {Chelli}\ \emph
  {et~al.}(2007{\natexlab{b}})\citenamefont {Chelli}, \citenamefont {Marsili},
  \citenamefont {Barducci},\ and\ \citenamefont
  {Procacci}}]{chelli2007numerical}%
  \BibitemOpen
  \bibfield  {author} {\bibinfo {author} {\bibfnamefont {Riccardo}\
  \bibnamefont {Chelli}}, \bibinfo {author} {\bibfnamefont {Simone}\
  \bibnamefont {Marsili}}, \bibinfo {author} {\bibfnamefont {Alessandro}\
  \bibnamefont {Barducci}}, \ and\ \bibinfo {author} {\bibfnamefont {Piero}\
  \bibnamefont {Procacci}},\ }\bibfield  {title} {\enquote {\bibinfo {title}
  {Numerical verification of the generalized {C}rooks nonequilibrium work
  theorem for non-hamiltonian molecular dynamics simulations},}\ }\href@noop {}
  {\bibfield  {journal} {\bibinfo  {journal} {The Journal of Chemical Physics}\
  }\textbf {\bibinfo {volume} {127}} (\bibinfo {year}
  {2007}{\natexlab{b}})}\BibitemShut {NoStop}%
\bibitem [{Note5()}]{Note5}%
  \BibitemOpen
  \bibinfo {note} {We also show that the Jarzynski equality becomes the staged
  Zwanzig formula for free-energy perturbation if the trajectory remains in
  equilibrium, and becomes the formula for thermodynamic integration if, in
  addition, the control parameters change in infinitesimal increments; related
  limiting forms were derived in~Ref.~\cite {jarzynski1997nonequilibrium}
  within the framework of Hamiltonian dynamics.}\BibitemShut {Stop}%
\bibitem [{\citenamefont {Brey}\ and\ \citenamefont
  {Casado}(1990)}]{brey1990generalized}%
  \BibitemOpen
  \bibfield  {author} {\bibinfo {author} {\bibfnamefont {JJ}~\bibnamefont
  {Brey}}\ and\ \bibinfo {author} {\bibfnamefont {J}~\bibnamefont {Casado}},\
  }\bibfield  {title} {\enquote {\bibinfo {title} {Generalized langevin
  equations with time-dependent temperature},}\ }\href@noop {} {\bibfield
  {journal} {\bibinfo  {journal} {Journal of statistical physics}\ }\textbf
  {\bibinfo {volume} {61}},\ \bibinfo {pages} {713--722} (\bibinfo {year}
  {1990})}\BibitemShut {NoStop}%
\bibitem [{\citenamefont {Dellago}\ \emph {et~al.}(1998)\citenamefont
  {Dellago}, \citenamefont {Bolhuis},\ and\ \citenamefont
  {Chandler}}]{dellago1998efficient}%
  \BibitemOpen
  \bibfield  {author} {\bibinfo {author} {\bibfnamefont {Christoph}\
  \bibnamefont {Dellago}}, \bibinfo {author} {\bibfnamefont {Peter~G}\
  \bibnamefont {Bolhuis}}, \ and\ \bibinfo {author} {\bibfnamefont {David}\
  \bibnamefont {Chandler}},\ }\bibfield  {title} {\enquote {\bibinfo {title}
  {Efficient transition path sampling: Application to {L}ennard-{J}ones cluster
  rearrangements},}\ }\href@noop {} {\bibfield  {journal} {\bibinfo  {journal}
  {The Journal of chemical physics}\ }\textbf {\bibinfo {volume} {108}},\
  \bibinfo {pages} {9236--9245} (\bibinfo {year} {1998})}\BibitemShut {NoStop}%
\bibitem [{\citenamefont {Whitelam}(2023)}]{whitelam2023demon}%
  \BibitemOpen
  \bibfield  {author} {\bibinfo {author} {\bibfnamefont {Stephen}\ \bibnamefont
  {Whitelam}},\ }\bibfield  {title} {\enquote {\bibinfo {title} {Demon in the
  machine: learning to extract work and absorb entropy from fluctuating
  nanosystems},}\ }\href@noop {} {\bibfield  {journal} {\bibinfo  {journal}
  {Physical Review X}\ }\textbf {\bibinfo {volume} {13}},\ \bibinfo {pages}
  {021005} (\bibinfo {year} {2023})}\BibitemShut {NoStop}%
\bibitem [{Note6()}]{Note6}%
  \BibitemOpen
  \bibinfo {note} {We could change $J$ at fixed $\beta $ in order to mimic
  temperature variation, but our model study is carried out to represent an
  experiment in which we cannot change the microscopic parameters of a
  material, while we can use temperature as a control parameter to take us
  across a phase boundary.}\BibitemShut {Stop}%
\bibitem [{tra()}]{trap_github}%
  \BibitemOpen
  \href@noop {} {}\bibinfo {howpublished}
  {\url{https://github.com/swhitelam/trap}}\BibitemShut {NoStop}%
\bibitem [{dem()}]{demon_github}%
  \BibitemOpen
  \href@noop {} {}\bibinfo {howpublished}
  {\url{https://github.com/swhitelam/demon}}\BibitemShut {NoStop}%
\bibitem [{\citenamefont {Evans}\ and\ \citenamefont
  {Searles}(2002)}]{evans2002fluctuation}%
  \BibitemOpen
  \bibfield  {author} {\bibinfo {author} {\bibfnamefont {Denis~J}\ \bibnamefont
  {Evans}}\ and\ \bibinfo {author} {\bibfnamefont {Debra~J}\ \bibnamefont
  {Searles}},\ }\bibfield  {title} {\enquote {\bibinfo {title} {The fluctuation
  theorem},}\ }\href@noop {} {\bibfield  {journal} {\bibinfo  {journal}
  {Advances in Physics}\ }\textbf {\bibinfo {volume} {51}},\ \bibinfo {pages}
  {1529--1585} (\bibinfo {year} {2002})}\BibitemShut {NoStop}%
\bibitem [{\citenamefont {Seifert}(2012)}]{seifert2012stochastic}%
  \BibitemOpen
  \bibfield  {author} {\bibinfo {author} {\bibfnamefont {Udo}\ \bibnamefont
  {Seifert}},\ }\bibfield  {title} {\enquote {\bibinfo {title} {Stochastic
  thermodynamics, fluctuation theorems and molecular machines},}\ }\href@noop
  {} {\bibfield  {journal} {\bibinfo  {journal} {Reports on progress in
  physics}\ }\textbf {\bibinfo {volume} {75}},\ \bibinfo {pages} {126001}
  (\bibinfo {year} {2012})}\BibitemShut {NoStop}%
\bibitem [{\citenamefont {Pohorille}\ \emph {et~al.}(2010)\citenamefont
  {Pohorille}, \citenamefont {Jarzynski},\ and\ \citenamefont
  {Chipot}}]{pohorille2010good}%
  \BibitemOpen
  \bibfield  {author} {\bibinfo {author} {\bibfnamefont {Andrew}\ \bibnamefont
  {Pohorille}}, \bibinfo {author} {\bibfnamefont {Christopher}\ \bibnamefont
  {Jarzynski}}, \ and\ \bibinfo {author} {\bibfnamefont {Christophe}\
  \bibnamefont {Chipot}},\ }\bibfield  {title} {\enquote {\bibinfo {title}
  {Good practices in free-energy calculations},}\ }\href@noop {} {\bibfield
  {journal} {\bibinfo  {journal} {The Journal of Physical Chemistry B}\
  }\textbf {\bibinfo {volume} {114}},\ \bibinfo {pages} {10235--10253}
  (\bibinfo {year} {2010})}\BibitemShut {NoStop}%
\bibitem [{\citenamefont {Zwanzig}(1954)}]{zwanzig1954high}%
  \BibitemOpen
  \bibfield  {author} {\bibinfo {author} {\bibfnamefont {Robert~W}\
  \bibnamefont {Zwanzig}},\ }\bibfield  {title} {\enquote {\bibinfo {title}
  {High-temperature equation of state by a perturbation method. {I.} {N}onpolar
  gases},}\ }\href@noop {} {\bibfield  {journal} {\bibinfo  {journal} {The
  Journal of Chemical Physics}\ }\textbf {\bibinfo {volume} {22}},\ \bibinfo
  {pages} {1420--1426} (\bibinfo {year} {1954})}\BibitemShut {NoStop}%
\bibitem [{\citenamefont {Kirkwood}(1935)}]{kirkwood1935statistical}%
  \BibitemOpen
  \bibfield  {author} {\bibinfo {author} {\bibfnamefont {John~G}\ \bibnamefont
  {Kirkwood}},\ }\bibfield  {title} {\enquote {\bibinfo {title} {Statistical
  mechanics of fluid mixtures},}\ }\href@noop {} {\bibfield  {journal}
  {\bibinfo  {journal} {The Journal of Chemical Physics}\ }\textbf {\bibinfo
  {volume} {3}},\ \bibinfo {pages} {300--313} (\bibinfo {year}
  {1935})}\BibitemShut {NoStop}%
\bibitem [{\citenamefont {Frenkel}\ and\ \citenamefont
  {Smit}(2001)}]{frenkel2001understanding}%
  \BibitemOpen
  \bibfield  {author} {\bibinfo {author} {\bibfnamefont {Daan}\ \bibnamefont
  {Frenkel}}\ and\ \bibinfo {author} {\bibfnamefont {Berend}\ \bibnamefont
  {Smit}},\ }\href@noop {} {\emph {\bibinfo {title} {Understanding molecular
  simulation: from algorithms to applications}}},\ Vol.~\bibinfo {volume} {1}\
  (\bibinfo  {publisher} {Academic Press},\ \bibinfo {year} {2001})\BibitemShut
  {NoStop}%
\end{thebibliography}
\end{document}